\documentclass[12pt]{article}
\usepackage{amssymb}
\usepackage{amsmath}
\usepackage{graphicx}
\begin{document}
\graphicspath{{./Figures/}{.}}
\newtheorem{theorem}{Theorem}[section]
\newtheorem{lemma}{Lemma}[section]
\def\jandk{Jones $\&$ Kompala}
\def\sacc{S.~cerevisiae}

\title{Discontinuity Induced Bifurcations in a Model of
{\em Saccharomyces cerevisiae}.}
\author{D.J.W.~Simpson$^1$, D.S.~Kompala$^2$, J.D.~Meiss$^1$\thanks{
Supported from the NSF under grant DMS-0707659.
}
\vspace{2mm}\\
{\small 1.} Department of Applied Mathematics\\
University of Colorado\\
Boulder, CO 80309-0526\\
{\small 2.} Department of Chemical Engineering\\
University of Colorado\\
Boulder, CO 80309-0424}
\date{\today}
\maketitle

\begin{abstract}
We perform a bifurcation analysis of the mathematical model
of \jandk~[K.D.~Jones and D.S.~Kompala,
Cybernetic model of the growth dynamics
of {\em Saccharomyces cerevisiae}
in batch and continuous cultures,
{\em J. Biotech.},
71:105-131, 1999].
Stable oscillations arise via Andronov-Hopf bifurcations
and exist for intermediate values of the dilution rate
as has been noted from experiments previously.
A variety of discontinuity induced bifurcations arise
from a lack of global differentiability.
We identify and classify discontinuous bifurcations
including several codimension-two scenarios.
Bifurcation diagrams are explained by a general
unfolding of these singularities.
\end{abstract}

\section{Introduction}
\label{sec:INTRO}
\setcounter{equation}{0}

Yeasts are single-celled fungi of which more than one
thousand different species have been identified. The most commonly used
yeast is {\em Saccharomyces cerevisiae}
which has been utilized for the production of 
bread, wine and beer for thousands of years.
Biologists in a wide variety of fields use
{\em \sacc} as a model organism.

A common experimental method for observing biochemical processes
involved in yeast growth is
that of continuous cultivation in a chemostat \cite{BaOl86}.
The cell growth takes place in a vessel that is continuously stirred.
A nutrient containing fluid is pumped into the vessel
and cell culture flows out of
the vessel at the same rate, ensuring that the volume of
culture in the reaction vessel remains constant. The rate of flow in
and out divided by culture volume
is called the {\em dilution rate}. Quantities such as 
concentrations of chemicals can be measured in a variety of ways,
see \cite{PoMa88,SaKu92} for methods used in {\em \sacc}
experiments.

As a continuous culture experiment is carried out, it is common for the
system to reach a steady state. At the steady state, the rate of
cell division in the culture is equal to the dilution rate. However
experimentalists in the late 1960's, \cite{FiVo68}, observed 
that instead of settling to a steady state, continuous culture
experiments of {\em \sacc} could in some cases produce stable oscillations.
Von Meyenburg, \cite{vo73}, discovered in
subsequent experiments that these oscillations only occur in
an intermediate range of values of the dilution rate (between about
$0.08h^{-1}$ and $0.22h^{-1}$). Much work has since been done to
understand the cause of such oscillations,
see for instance \cite{Ri03,He03,DaHy01}.

{\em \sacc} has three metabolic pathways for glucose:
fermentation, ethanol oxidation and glucose oxidation.
The model of \jandk~\cite{JoKo99}
hypothesizes that the competing metabolic pathways 
of the growing yeast cells create feedback responses that
produce stable oscillations.
It assumes that 
micro-organisms will utilize the available substrates 
in a manner that maximizes their growth rate at all times.
To enforce this optimization
a ``maximum function'' is introduced in the model equations;
as a result, the model is an example of a
piecewise-smooth, continuous dynamical system.

Piecewise-smooth systems are characterized by the presence of codimension-one
phase-space boundaries, called {\em switching manifolds},
on which smoothness is lost.
Such systems have been utilized in diverse fields to model
non-smooth behavior, for example
vibro-impacting systems and systems with friction
\cite{WiDe00,Br99,LeNi04},
switching in electrical circuits
\cite{BaVe01,ZhMo03,Ts03},
economics
\cite{PuSu06,MoLa07}
and biology and physiology,
\cite{Ro70,KeSn98}.

The interaction of invariant sets with switching manifolds often produces bifurcations
that are forbidden in smooth systems.
For instance, though period-doubling cascades are a common
mechanism for the transition to chaos in smooth systems,
in piecewise-smooth systems periodic orbits may undergo
direct transitions to chaos \cite{DiBu08,PiVi04}.
These so-called {\em discontinuity induced bifurcations}
can be nonsmooth analogues of familiar smooth bifurcations
or can be novel bifurcations and unique to piecewise-smooth systems.
A bifurcation in the latter category that is simple in appearance
(for example the transition from a stable period-1 solution to a stable period-3
solution in a piecewise-smooth map)
often corresponds to a combination of or countable sequence of smooth bifurcations.
In this situation, arguably, the piecewise-smooth system
describes the dynamics more succinctly than any smooth system is able to.
Alternatively bifurcations in piecewise-smooth systems may be extremely complicated,
see for instance \cite{DiBu08,LeNi04,ZhMo03} and references within.

A piecewise-smooth, continuous system is one that is everywhere continuous
but nondifferentiable on switching manifolds.
In such a system, the collision of a mathematical equilibrium
(i.e.~steady state, abbreviated to equilibrium throughout this paper)
with a switching manifold
may give rise to a {\em discontinuous bifurcation}.
As the equilibrium crosses the switching manifold,
its associated eigenvalues generically change discontinuously.
This may produce a stability change and bifurcation.
In two-dimensional systems, all codimension-one discontinuous
bifurcations have been classified \cite{FrPo98},
but in higher dimensions
there are more allowable geometries
and no general classification is known.
See for instance \cite{FrPo07,CaFr06} for recent
investigations into three-dimensional systems.

In this paper we present an analysis of discontinuity induced bifurcations
in the eight-dimensional {\em \sacc}~model of \jandk~\cite{JoKo99}.
The model equations are stated in \S\ref{sec:MODEL}.
In \S\ref{sec:BIFSET} we illustrate a two-parameter bifurcation set
indicating parameter values at which stable oscillations occur.
The bifurcation set also shows curves
corresponding to codimension-one discontinuous bifurcations.
These bifurcations are analogous to saddle-node
and Andronov-Hopf bifurcations in smooth systems.
Bifurcations relating to stable oscillations are
described in \S\ref{sec:OSCIL}.
We observe period-adding sequences over small regions in parameter space.
In \S\ref{sec:CODIM2} we provide rigorous unfoldings
of codimension-two scenarios seen in the bifurcation set
from a general viewpoint.
Finally \S\ref{sec:CONCL} presents conclusions.

\section{A model of the growth of {\em Saccharomyces cerevisiae}}
\label{sec:MODEL}
\setcounter{equation}{0}

\jandk~\cite{JoKo99} give the following model equations:\\
\begin{equation}
\begin{split}
\frac{{\rm d}X}{{\rm d}t} & =
\left( \sum_i r_i v_i - D \right) X \;, \\
\frac{{\rm d}G}{{\rm d}t} & =
(G_0 - G)D - \left( \frac{r_1 v_1}{Y_1} +
\frac{r_3 v_3}{Y_3} \right)X - \phi_4
\left(
C \frac{{\rm d}X}{{\rm d}t} +
X \frac{{\rm d}C}{{\rm d}t} \right) \;, \\
\frac{{\rm d}E}{{\rm d}t} & =
-DE + \left( \phi_1 \frac{r_1 v_1}{Y_1}
- \frac{r_2 v_2}{Y_2} \right)X \;, \\
\frac{{\rm d}O}{{\rm d}t} & =
k_La(O^* - O) - \left( \phi_2 \frac{r_2 v_2}{Y_2} +
\phi_3 \frac{r_3 v_3}{Y_3} \right) X \;, \\
\frac{{\rm d}e_1}{{\rm d}t} & =
\alpha u_1 \frac{G}{K_1 + G} - 
\left( \sum_i r_i v_i + \beta \right) e_1 + \alpha^* \;, \\
\frac{{\rm d}e_2}{{\rm d}t} & =
\alpha u_2 \frac{E}{K_2 + E} - 
\left( \sum_i r_i v_i + \beta \right) e_2 + \alpha^* \;, \\
\frac{{\rm d}e_3}{{\rm d}t} & =
\alpha u_3 \frac{G}{K_3 + G} - 
\left( \sum_i r_i v_i + \beta \right) e_3 + \alpha^* \;, \\
\frac{{\rm d}C}{{\rm d}t} & =
\gamma_3 r_3 v_3 - (\gamma_1 r_1 v_1 + \gamma_2 r_2 v_2)C
- \left( \sum_i r_i v_i \right) C \;,
\label{eq:ALLdot}
\end{split}
\end{equation}where $X, G, E$ and $O$ represent the concentrations of cell mass,
glucose, ethanol (in gL$^{-1}$) and dissolved oxygen
(in mgL$^{-1}$) in the culture volume, respectively.
Each $e_i$ represents the
intracellular mass fraction of a
key enzyme in the $i^{\rm th}$
metabolic pathway. $C$ represents the intracellular carbohydrate
mass fraction. The subscripts $1,2$ and $3$ correspond to the
three pathways: fermentation, ethanol oxidation and glucose
oxidation, respectively; $r_i$ represents the growth rate
on each pathway. Formulas for the growth rates and other functions
are given in Appendix \ref{sec:PARAM}. Details concerning the derivation
of the model are found in \cite{JoKo99} and expanded in the 
M.S.~thesis of Jones \cite{Jo95}.

In particular, the equation
\begin{equation}
v_i = \frac{r_i}{\max(r_1,r_2,r_3)} \;,
\label{eq:v_i}
\end{equation}
is introduced to model the regulation of enzyme activity
by numerous biochemical mechanisms.
Each $v_i$ is a smooth function except
at points where the two largest $r_i$ are equal;
at these points each $v_i$ has discontinuous derivatives with
respect to some of the variables. All eight differential
equations have at least one term containing a $v_i$ and thus display
the same lack of smoothness properties.
Therefore the model is a piecewise-smooth, continuous system.
Our goal is to understand
the effects that this nonsmoothness has on the resulting dynamics.

This paper focuses on variations in values of two parameters, namely
the dilution rate, $D$ (in h$^{-1}$),
and the dissolved oxygen mass transfer coefficient,
$k_La$ (in h$^{-1}$).
Values for all other parameters are given in Appendix \ref{sec:PARAM}
and were kept constant throughout the investigations.

A key property of the system (\ref{eq:ALLdot}), 
is that the positive hyper-octant is forward invariant.
That is, if the values of all variables are initially positive,
they will remain positive for all time.
This is, of course,
a property that is required for any sensible model,
since negative values of the variables are not physical.
Furthermore, within the positive hyper-octant
all trajectories are bounded forward in time.
In other words, solutions always approach some attracting set
which could be an equilibrium, a periodic orbit or a more
complicated and possibly chaotic attractor.

\section{A bifurcation set}
\label{sec:BIFSET}
\setcounter{equation}{0}

\begin{figure}[h]
\begin{center}
\includegraphics[width=13.2cm,height=11cm]{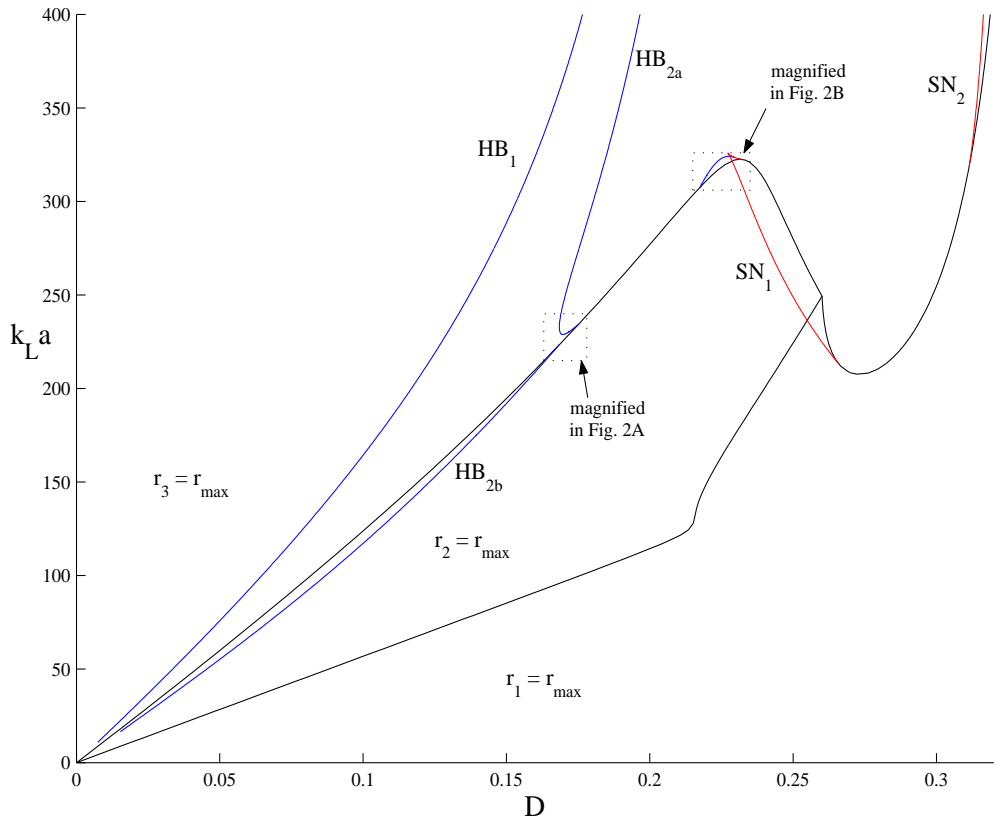}
\caption{A bifurcation set for the system
(\ref{eq:ALLdot}).
HB - Hopf bifurcation,
SN - saddle-node bifurcation.
The parameter space is divided into three regions
within each of which a different metabolic pathway is preferred
at equilibrium.
\label{fig:bifset}}
\end{center}
\end{figure}

In this section we describe a numerically computed bifurcation set
for the system (\ref{eq:ALLdot}), see Fig.~\ref{fig:bifset}.
We find a single, physically meaningful equilibrium (steady-state)
except in small windows of parameter space between saddle-node bifurcations.
For small values of $D$, this equilibrium is stable.
If we fix the value of $k_La$ and increase $D$,
the first bifurcation encountered is an
Andronov-Hopf bifurcation (labelled ${\rm HB}_1$ in Fig.~\ref{fig:bifset}).
Slightly to the right of ${\rm HB}_1$ the equilibrium 
is unstable and solutions approach a periodic orbit or
complicated attractor (see \S\ref{sec:OSCIL}).
As the value of $D$ is increased further a second Hopf bifurcation
(${\rm HB}_{2a}$ or ${\rm HB}_{2b}$)
is encountered that restores stability to the equilibrium
(unless $k_La \approx 230$, see below).

\begin{figure}[h]
\begin{center}
\setlength{\unitlength}{1cm}
\begin{picture}(14.7,6)
\put(0,0){\includegraphics[width=7.2cm,height=6cm]{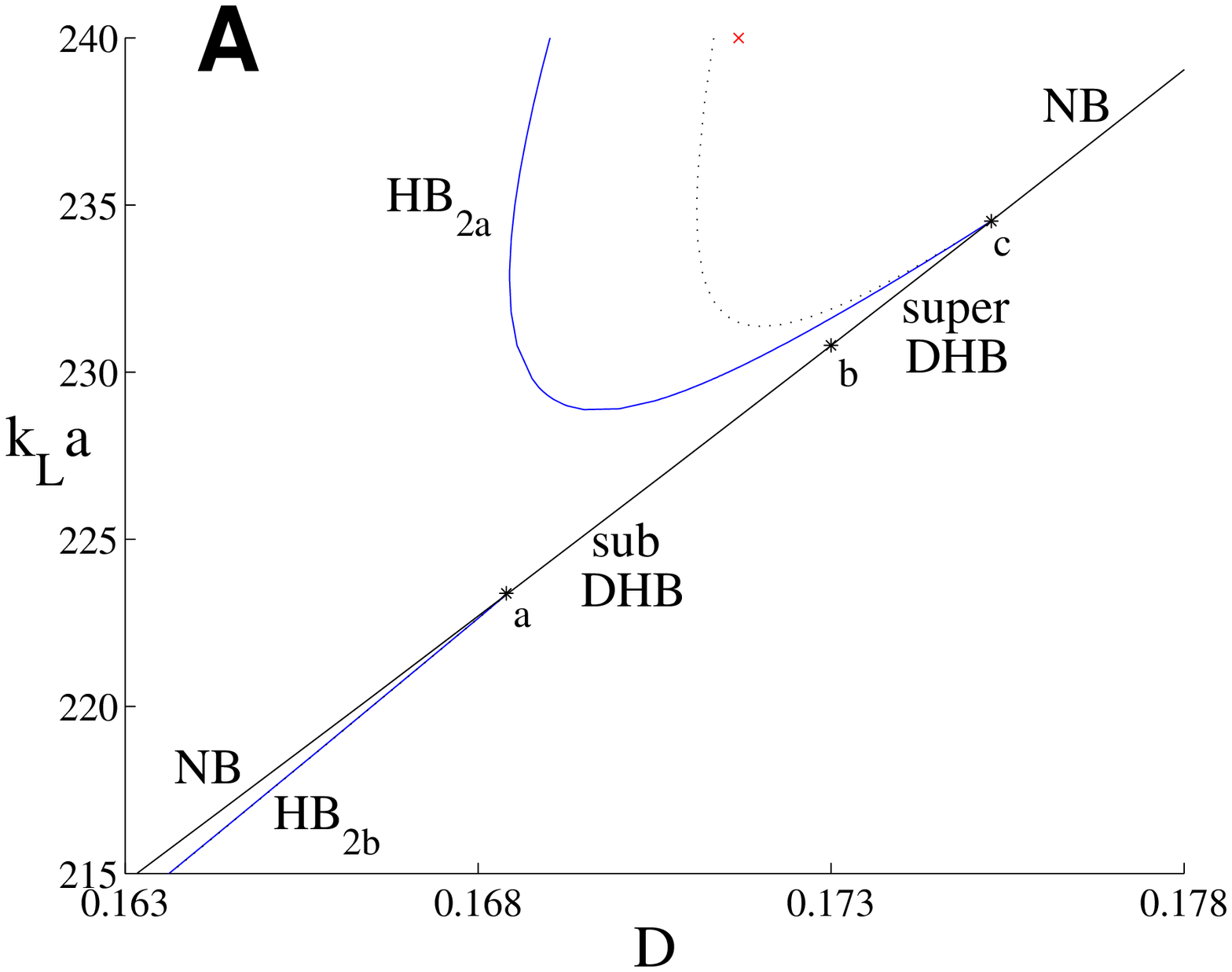}}
\put(7.5,0){\includegraphics[width=7.2cm,height=6cm]{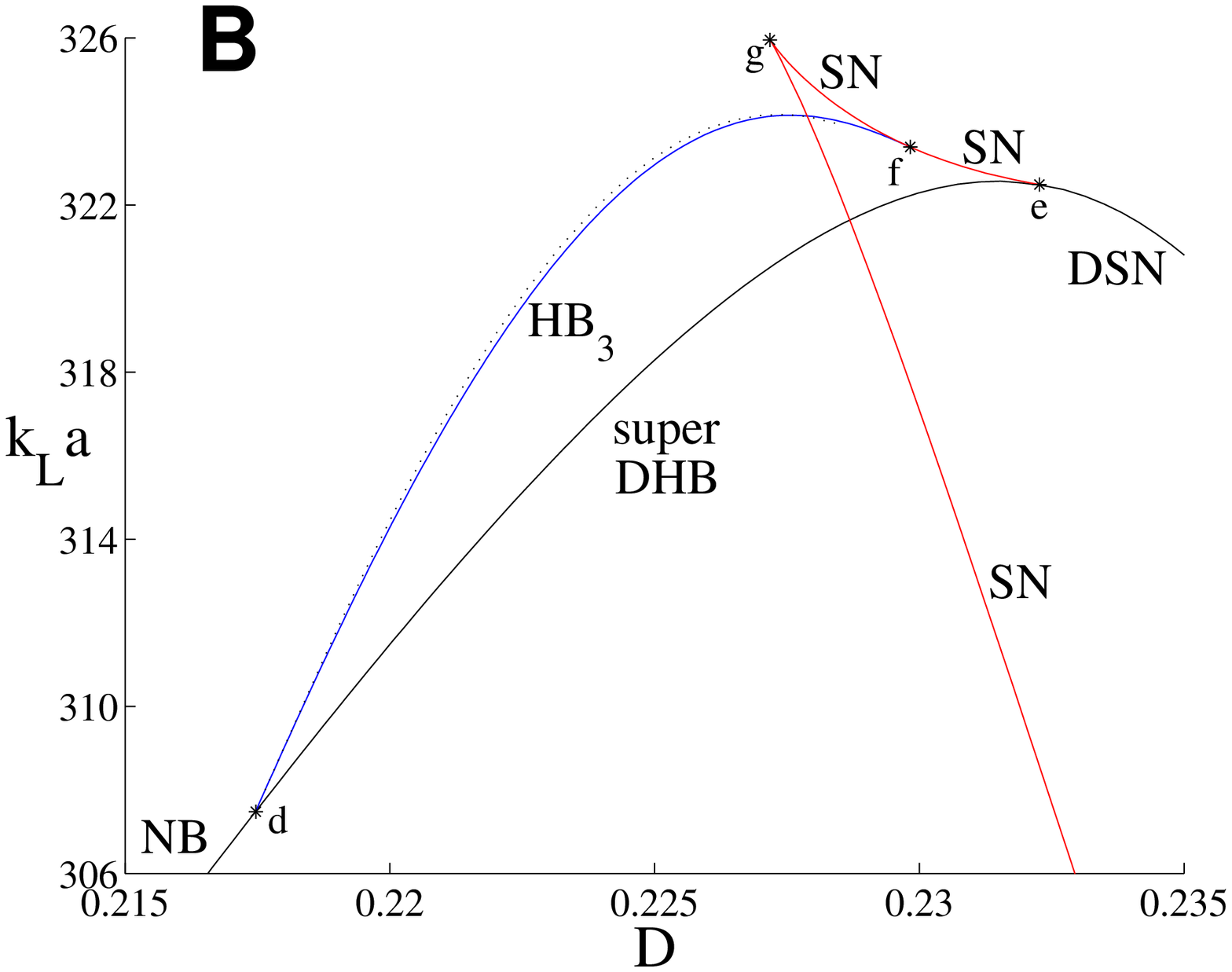}}
\end{picture}
\end{center}
\caption{Magnified views of Fig.~\ref{fig:bifset}.
HB - Hopf bifurcation, SN - saddle-node bifurcation.
The curves of discontinuity are labelled by 
their corresponding bifurcations: NB - no bifurcation,
DHB - discontinuous Hopf bifurcation (with criticality indicated),
DSN - discontinuous saddle-node bifurcation.
The dotted curves correspond to the grazing of a Hopf cycle
with a switching manifold.
There are three such curves, these emanate from the points
(a), (c) and (d) (note, the curve that emanates from (a) is barely
distinguishable from ${\rm HB}_{2b}$). 
In panel A, one point on a locus of saddle-node bifurcations of a Hopf cycle
is shown with a red cross.
\label{fig:bsmag}}
\end{figure}

Recall, the system is piecewise-smooth, continuous
as a result of a maximum function
in the rate coefficient, (\ref{eq:v_i}).
Since the switching manifold is codimension-one,
it is a codimension-one phenomenon for the equilibrium
to lie precisely on a switching manifold.
Though an analytical formula for the equilibrium seems difficult to obtain,
we have been able to numerically compute curves in
two-dimensional parameter space along which this
codimension-one situation occurs -
the black curves in Fig.~\ref{fig:bifset}.
We refer to these as {\em curves of discontinuity}.
The curves of discontinuity divide parameter space into three regions where
one of the $r_i$ is larger than the other two,
at the equilibrium.
They may also correspond to discontinuous bifurcations,
as described below.
Physically, crossing a curve of discontinuity
corresponds to a change in the preferred metabolic pathway
at equilibrium.

Fig.~\ref{fig:bsmag} shows an enlargement
of Fig.~\ref{fig:bifset} near two points on a curve of discontinuity.
In panel A, along the curve of discontinuity,
below the point (a) and above the point (c),
there is no bifurcation.
Between (a) and (c), numerically we have observed that
a periodic orbit is created when the equilibrium crosses the switching manifold.
Between (a) and (b), the orbit is unstable and
emanates to the right of the curve of discontinuity.
Between (b) and (c) the orbit is stable and emanates to the left.
We refer to these bifurcations as subcritical and supercritical
{\em discontinuous Hopf bifurcations}, respectively.
The codimension-two point (b), is akin to a Hopf bifurcation
at which the constant determining criticality vanishes.
We expect that a locus of saddle-node bifurcations will emanate
from this point, in manner similar to that in smooth systems.

Two of the loci of smooth Hopf bifurcations,
${\rm HB}_{2a}$ and ${\rm HB}_{2b}$,
collide with the curve of discontinuity at (a) and (c).
Near these points these Hopf bifurcations are subcritical.
Unstable periodic orbits emanate from the Hopf bifurcations
and are initially of sufficiently small amplitude to
not intersect a switching manifold.
However, as we move away from the Hopf bifurcations,
the amplitude of the Hopf cycles grow
and they graze the switching manifold along the dotted curves
in Fig.~\ref{fig:bsmag}.
No bifurcation occurs at the grazing because the system
is continuous \cite{DiBu01}.
As we will show in Theorem \ref{th:c2hb},
the grazing curves intersect the Hopf loci tangentially.
The unstable cycles persist beyond grazing
until they collide with a stable cycle in a saddle-node bifurcation.
Loci of saddle-node bifurcations of periodic orbits
are not shown in the figures because we have not been able
to accurately numerically compute more than a single point
(when $k_La = 240$, see Fig.~\ref{fig:bsmag}A) on the curves
due to the stiffness, non-smoothness and high dimensionality
of the system (\ref{eq:ALLdot}).
We expect one such curve to emanate from (c)
and lie extremely close to the upper grazing curve
as has recently been shown for two-dimensional systems \cite{SiMe08}.

Fig.~\ref{fig:bsmag}B, shows a second magnification of Fig.~\ref{fig:bifset}
near $D = 0.225$ and $k_La = 320$.
Loci of Hopf bifurcations and saddle-node bifurcations
of the equilibrium have endpoints at (d) and (e) that lie on a curve of discontinuity.
We will show in \S\ref{sec:CODIM2} that
bifurcations and dynamical behavior in neighborhoods of (d) and (e)
are predicted by Theorems \ref{th:c2hb} and \ref{th:c2sn}, respectively.
To the left of the point (d), no bifurcation occurs along the curve of discontinuity.
To the right of (e), points on the curve of discontinuity act as saddle-node bifurcations,
hence we refer to these as {\em discontinuous saddle-node bifurcations}.
From the point (d) to very close to (e),
the curve of discontinuity
corresponds to a supercritical discontinuous Hopf bifurcation.
A Takens-Bogdanov bifurcation occurs at (f) where the Hopf locus,
${\rm HB}_3$, terminates
at the saddle-node locus, and the point (g)
corresponds to a cusp bifurcation \cite{Ku04}.
Near the points (e) and (f) we believe there are a variety of additional
bifurcations that we have not yet identified.
The system exhibits stable oscillations in the region between the
smooth Hopf bifurcations and discontinuous Hopf bifurcations,
but to our knowledge, oscillations in this parameter region have
not been observed experimentally.

\section{Simple and complicated stable oscillations}
\label{sec:OSCIL}
\setcounter{equation}{0}

\begin{figure}[b!]
\begin{center}
\includegraphics[width=13.2cm,height=11cm]{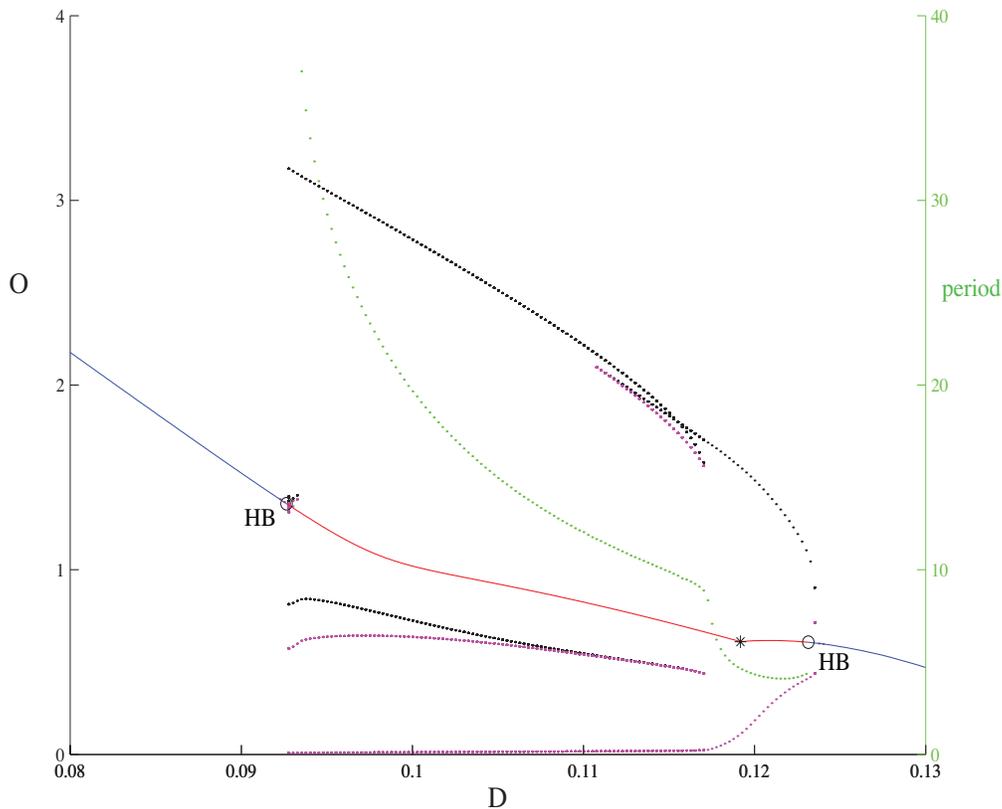}
\caption{A bifurcation diagram of the system
(\ref{eq:ALLdot}) when $k_La = 150$.
The equilibrium is colored blue when stable and red otherwise.
Black [magenta] dots correspond to local maxima [minima] $O$ that
stable oscillations obtain.
Two Hopf bifurcations are indicated by circles.
The asterisk corresponds to where $r_1 < r_2 = r_3$ at equilibrium.
The period of these oscillations is also indicated.
\label{fig:bif150}}
\end{center}
\end{figure}

\begin{figure}[h]
\begin{center}
\includegraphics[width=13.2cm,height=11cm]{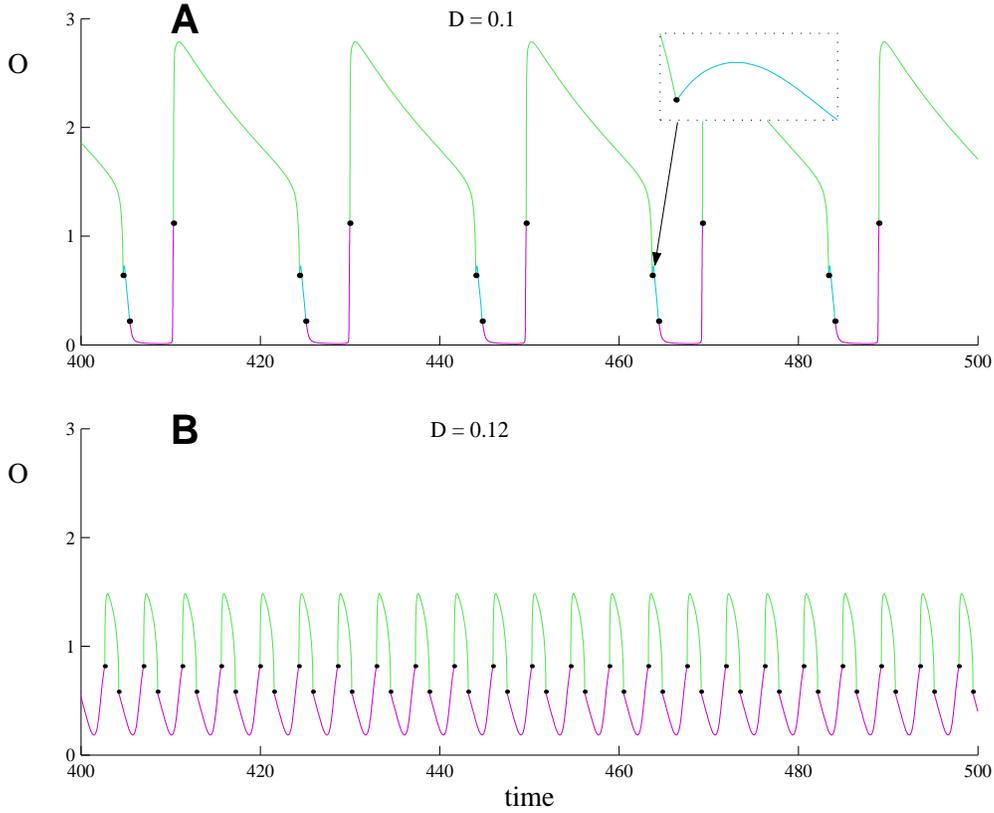}
\caption{Time series after transients have decayed
when $k_La = 150$, $D = 0.1$ in panel A and $D = 0.12$ in panel B. 
The solution is colored
cyan when $r_1 = r_{\rm max}$,
magenta when $r_2 = r_{\rm max}$ and
green when $r_3 = r_{\rm max}$
\label{fig:ts150}}
\end{center}
\end{figure}

The experimentally observed oscillations correspond to the region
between ${\rm HB}_1$ and ${\rm HB}_2$ in Fig.~\ref{fig:bifset}.
In this section, we will discuss the dynamics in this
region in more detail.
Fig.~\ref{fig:bif150}, shows a one parameter bifurcation diagram
of the system (\ref{eq:ALLdot})
when $k_La = 150$.
The black [magenta] dots represent local maxima [minima] $O$
on the stable oscillating cycle.
For values of $D$ between about $0.0927$ and $0.1172$,
all three pathways are at some time preferred
over one period of the solution,
see Fig.~\ref{fig:ts150}.
In particular, very soon after the preferred pathway changes from glucose oxidation
to fermentation (green to cyan in Fig.~\ref{fig:ts150}),
the concentration of dissolved oxygen rebounds slightly
before continuing to decrease.
Thus local maxima appear below the equilibrium value in Fig.~\ref{fig:bif150}.
For larger values of $D$,
still to the left of the rightmost Hopf bifurcation,
fermentation is no longer a preferred pathway at any
point on the stable solution, and the lower local maximum is lost.
Also, the absolute maximum undergoes two cusp catastrophes at
$D \approx 0.111,~0.117$.


\begin{figure}[h]
\begin{center}
\includegraphics[width=13.2cm,height=11cm]{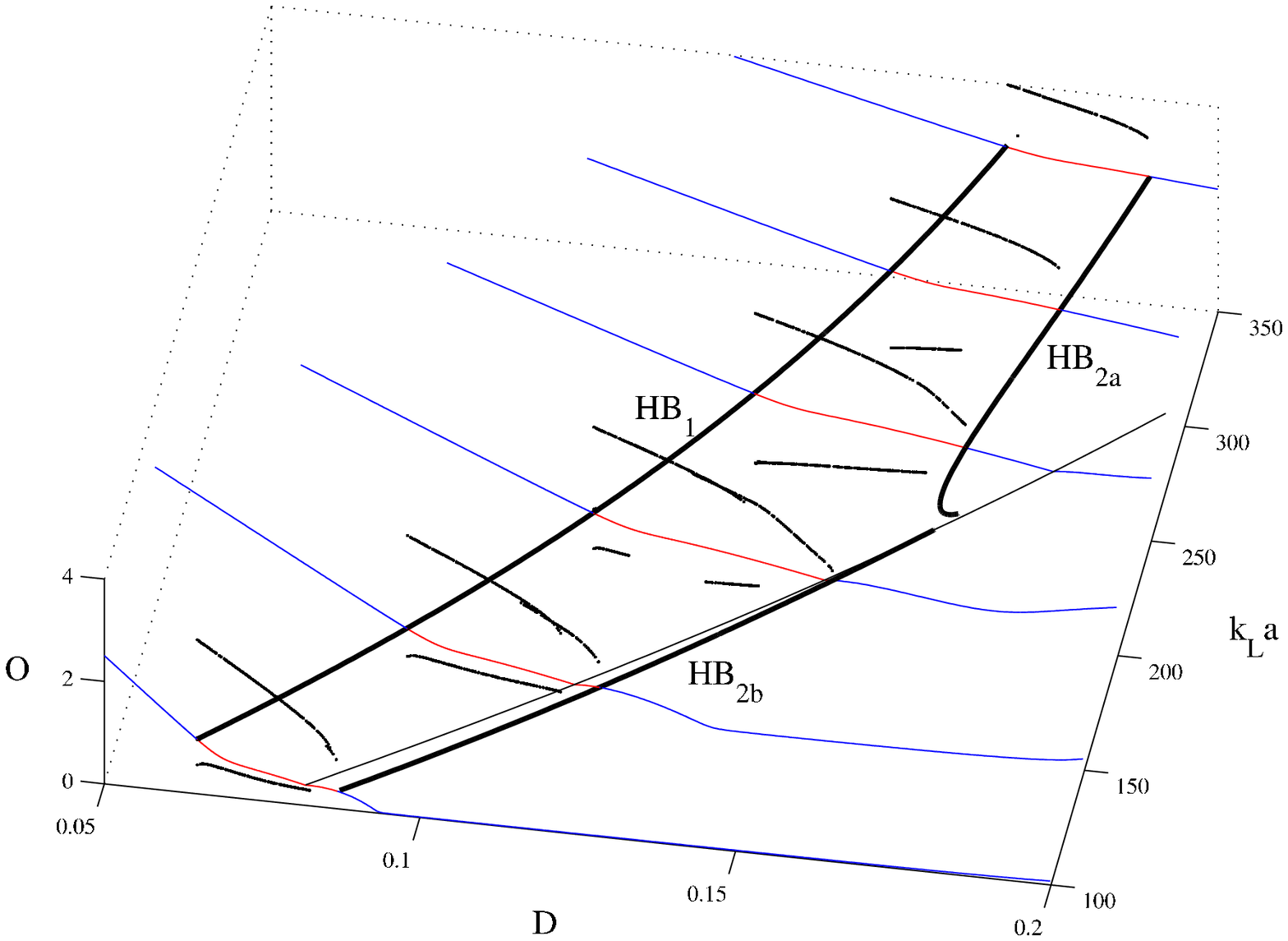}
\caption{Bifurcation diagrams of (\ref{eq:ALLdot})
at six different values of $k_La$ with the same color scheme
as Fig.~\ref{fig:bif150}.
A curve of discontinuity and Hopf loci
shown in Fig.~\ref{fig:bifset} are also included.
\label{fig:dots3d}}
\end{center}
\end{figure}

\begin{figure}[h]
\begin{center}
\setlength{\unitlength}{1cm}
\begin{picture}(13.2,11)
\put(0,0){\includegraphics[width=13.2cm,height=11cm]{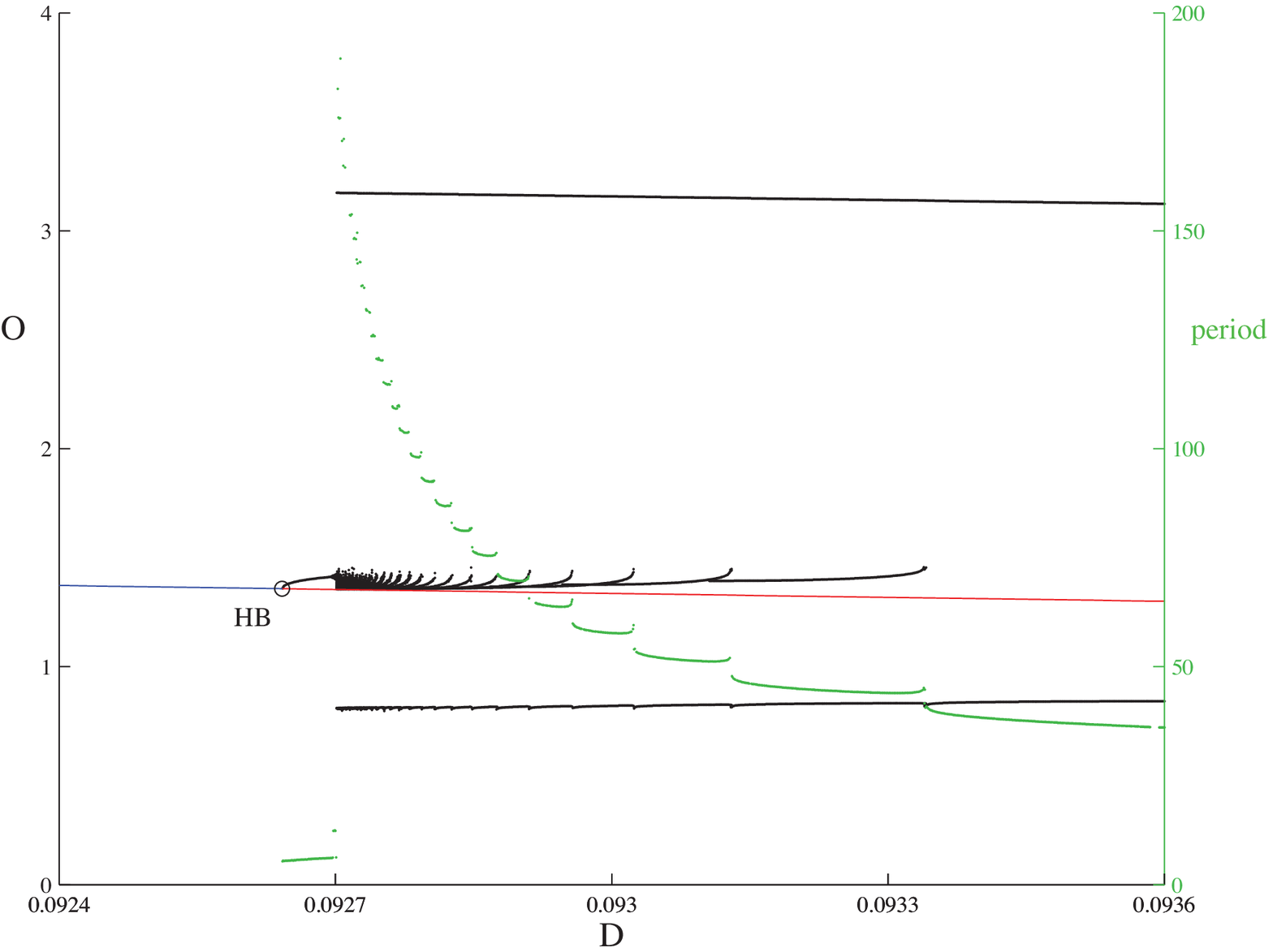}}
\end{picture}
\end{center}
\caption{A magnification of Fig.~\ref{fig:bif150},
near the leftmost Hopf bifurcation.
Local minimums are not shown.
\label{fig:dots150z}}
\end{figure}

Different values of $k_La$
yield similar bifurcation diagrams, we show a collection in Fig.~\ref{fig:dots3d}.
As a general rule there is a rapid change from a stable equilibrium
to a large amplitude orbit near the leftmost Hopf bifurcation
and as $D$ is increased the amplitude and period of the orbit decrease.

The behavior near the leftmost Hopf bifurcation is actually quite complex,
as indicated in Fig.~\ref{fig:dots150z},
which is a magnification of Fig.~\ref{fig:bif150}.
Here the Hopf bifurcation is supercritical giving rise to a stable orbit which
then undergoes a period-doubling cascade to chaos over an extremely small interval.
The first period-doubling occurs at $D \approx 0.092697$
and the solution appears chaotic by $D = 0.092700$.
At $D \approx 0.092701$ the attractor suddenly explodes in size
and the oscillation amplitude grows considerably.
As $D$ decreases toward this point we
observe a period-adding sequence.
Period-adding sequences are characterized by successive jumps in the
period in a manner that forms an approximately
arithmetic sequence.
Such sequences have been observed in models of many physical systems
\cite{PiVi04,YaLu06,PiTu04,TaMu03}.
To our knowledge period-adding sequences are not completely understood,
but seem to arise when periodic solutions interact with an invariant manifold
of a saddle-type equilibrium giving rise to
a Poincar\'{e} map that is piecewise-smooth and often discontinuous.
Period-adding in one-dimensional piecewise-smooth maps has been
the subject of recent research \cite{DiBu08,AvSc07,HaHo03}.
Dynamical behavior between period-adding windows
(intervals of the bifurcation parameter within with the period
undergoes no sudden change)
is determined by the types and order of
various local bifurcations.
In Fig.~\ref{fig:dots150z},
the period appears to go to infinity in the period-adding sequence.
Within the extremely small regions between windows,
we have identified period-doubling bifurcations
and complicated attractors although these attractors
deviate only slightly from the observed periodic orbits.

\section{Codimension-two discontinuous bifurcations}
\label{sec:CODIM2}
\setcounter{equation}{0}

\begin{figure}[h]
\begin{center}
\setlength{\unitlength}{1cm}
\begin{picture}(14.9,6)
\put(0,0){\includegraphics[width=7.2cm,height=6cm]{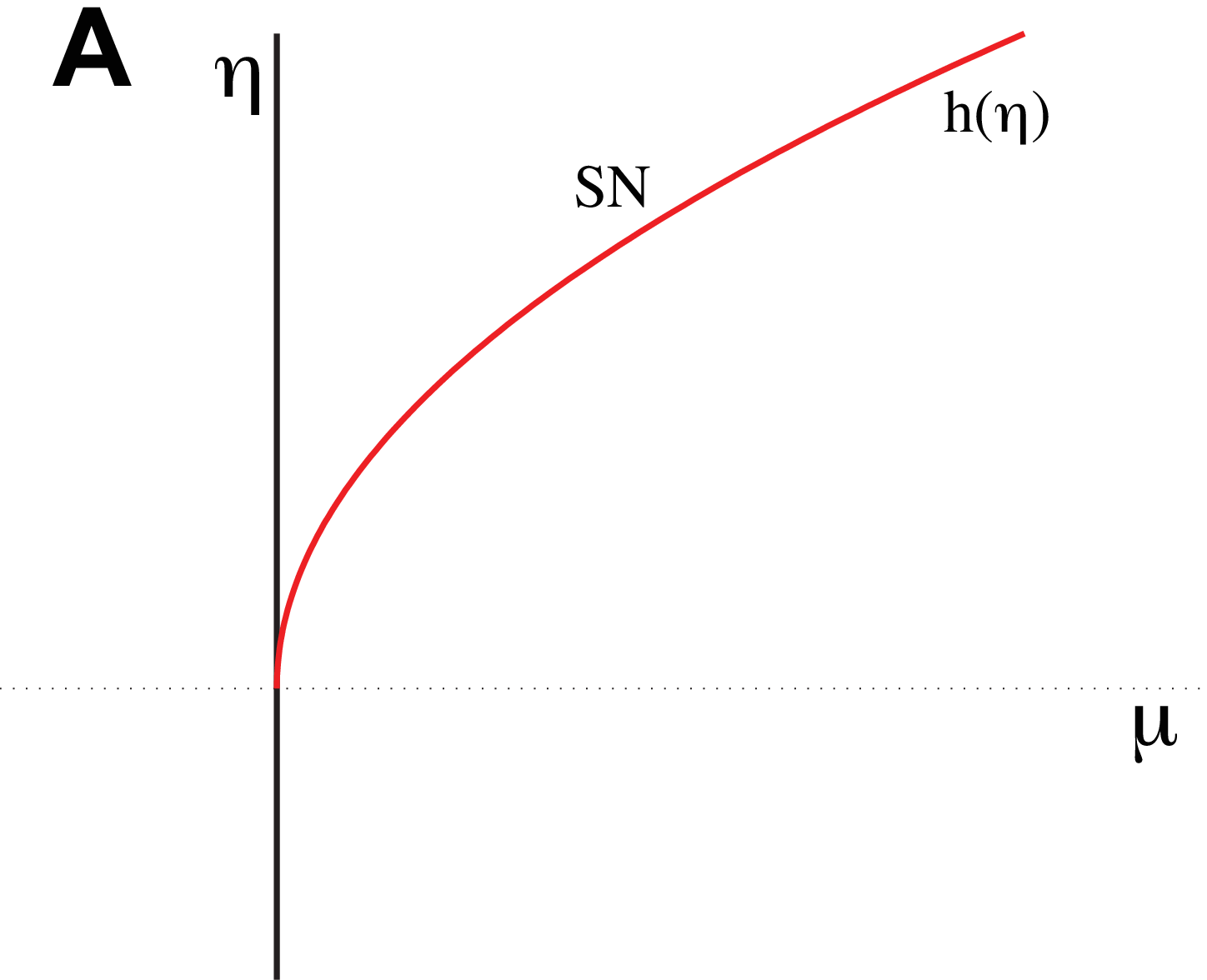}}
\put(7.7,0){\includegraphics[width=7.2cm,height=6cm]{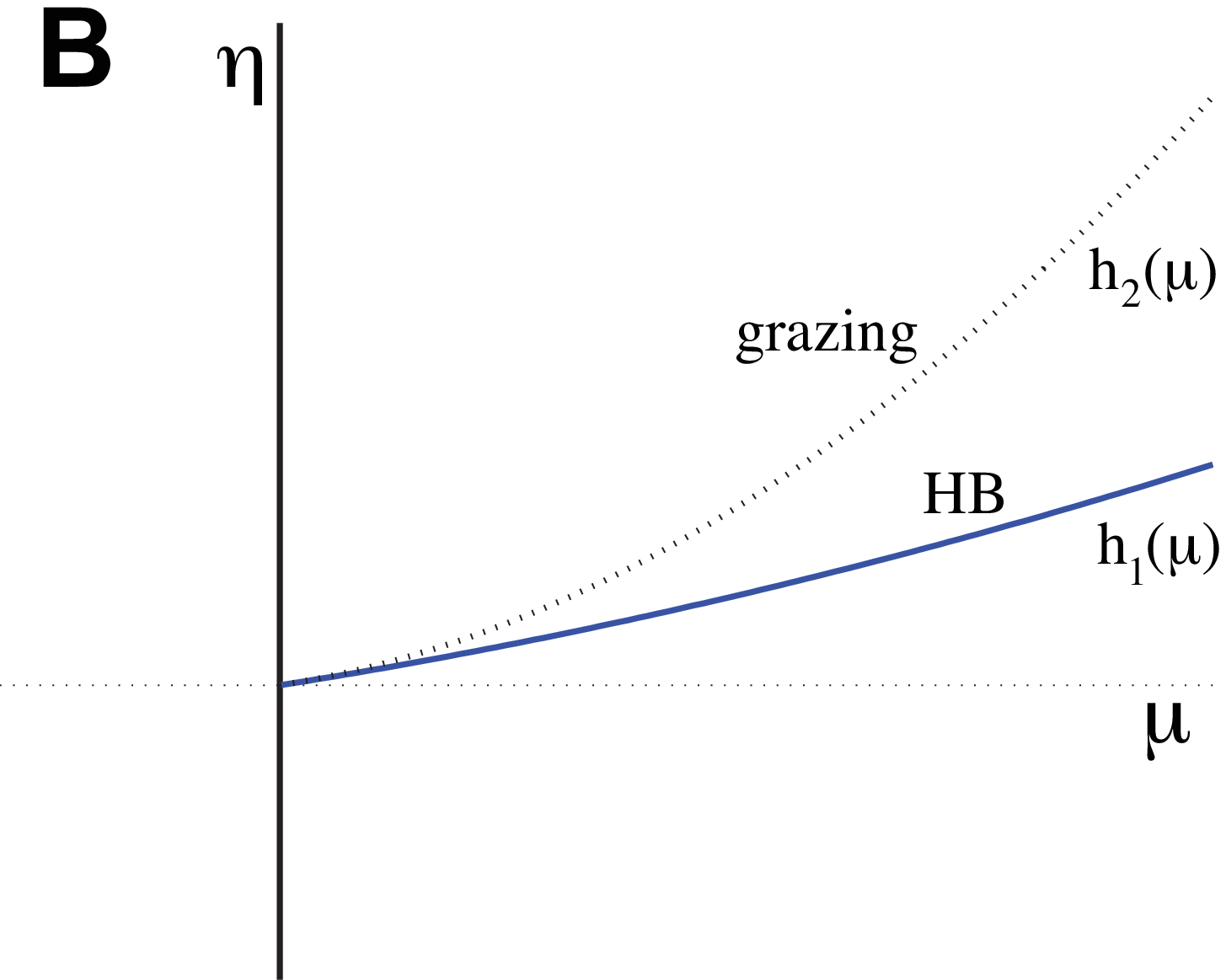}}
\end{picture}
\end{center}
\caption{
Unfoldings predicted by Theorems \ref{th:c2sn} and \ref{th:c2hb}.
SN - saddle-node bifurcation, HB - Hopf bifurcation.
Along the curve labelled ``grazing'', the Hopf cycle 
intersects the switching manifold at one point.
Along the $\eta$-axis,
an equilibrium lies on the switching manifold.
\label{fig:bsSch}}
\end{figure}

This section studies dynamics near two of the codimension-two,
discontinuous bifurcation scenarios that were identified in \S\ref{sec:BIFSET}.
Adopting a general viewpoint,
we will first unfold the simultaneous occurrence
of a saddle-node bifurcation and a
discontinuous bifurcation; our results are summarized in Fig.~\ref{fig:bsSch}A.
The tangency illustrated in this figure matches our numerically
computed bifurcation set, specifically point (e) of Fig.~\ref{fig:bsmag}B.
Secondly we will unfold the simultaneous occurrence of a Hopf bifurcation and
a discontinuous bifurcation, see Fig.~\ref{fig:bsSch}B.
This theoretical prediction also matches numerical results,
specifically the points (a), (c) and (d) of Fig.~\ref{fig:bsmag}.

The results of this section are presented
formally in Theorems \ref{th:c2sn} and \ref{th:c2hb},
proofs of which are given in Appendix \ref{sec:PROOFS}.
Proofs to Lemmas \ref{le:c2hb2d} and \ref{le:c2hb2dgen}
are described in \cite{SiMe08}.
Throughout this section we use arbitrary parameters $\mu$ and $\eta$
that do not relate to particular parameters of (\ref{eq:ALLdot}).
To clarify our notation, $O(i)$ [$o(i)$]
denotes terms that are order $i$ or greater [greater than order $i$]
in {\em every} variable and parameter on which the given expression depends.

In the neighborhood of a single switching manifold
an $N$-dimensional, piecewise-smooth, continuous
system may be written as
\begin{equation}
\dot{x} = \left\{ \begin{array}{lc} 
f^{(L)}(x;\mu,\eta), & H(x;\mu,\eta) \le 0 \\
f^{(R)}(x;\mu,\eta), & H(x;\mu,\eta) \ge 0
\end{array} \right. \;,
\label{eq:genFlow}
\end{equation}
where $f^{(L)},f^{(R)} : \mathbb{R}^N \times \mathbb{R}^2 \to \mathbb{R}^N$
are $C^k$ and we assume that
$H : \mathbb{R}^N \times \mathbb{R}^2 \to \mathbb{R}$
is sufficiently smooth (at least $C^4$).
The switching manifold is the parameter dependent set,
$\mathcal{S}_{\mu,\eta} = \{ x \in \mathbb{R}^N ~|~ H(x;\mu,\eta) = 0 \}$.
If, when $(x;\mu,\eta) = (0;0,0)$, we have $H = 0$ and 
$\nabla H \ne 0$
then locally $\mathcal{S}_{0,0}$ is a $(N-1)$-dimensional manifold
intersecting the origin.
Via coordinate transformations in a similar manner to those
given in \cite{DiBu01}, we may assume, to order $C^4$, that
$H$ is simply equal to $x_1$.
The higher order terms in $H$ do not affect our analysis below;
thus for simplicity, in what follows we will assume
$H$ is identically equal to $x_1$.
The switching manifold is then the plane $x_1 = 0$
and we will refer to
$f^{(L)}$ as the {\em left-half-system} and 
$f^{(R)}$ as the {\em right-half-system}.

We assume that when $\mu = \eta = 0$, the origin is an equilibrium.
Since the origin lies on the switching manifold and (\ref{eq:genFlow})
is continuous, it is an equilibrium of both left and right-half-systems.
We will assume
\begin{equation}
\det(D_x f^{(R)}(0;0,0)) \ne 0 \;;
\end{equation}
that is, for the right-half-system, zero is not an associated eigenvalue
of the origin when $\mu = \eta = 0$.
Then by the implicit function theorem
the right-half-system has an equilibrium, $x^{*(R)}(\mu,\eta)$,
with $x^{*(R)}(0,0) = 0$ and that depends upon the parameters
as a $C^k$ function in some neighborhood of the origin.
As is generically the case,
we may assume the distance the equilibrium is from the switching manifold
varies linearly with some linear combination of the parameters.
Without loss of generality we may assume $\mu$ is a suitable choice;
that is
\begin{equation}
\frac{\partial x^{*(R)}_1(0,0)}{\partial \mu} \ne 0 \;.
\label{eq:transvCond}
\end{equation}
In this case, the implicit function theorem implies
there is a $C^k$ function $\phi_1 : \mathbb{R} \to \mathbb{R}$
such that $x^{*(R)}(\phi_1(\eta),\eta) = 0$.
In other words when $\mu = \phi(\eta)$,
the equilibrium lies on the switching manifold.
By performing the nonlinear change of coordinates
\begin{eqnarray}
\mu & \mapsto & \mu - \phi_1(\eta) \;, \nonumber \\
x & \mapsto & x - x^{*(R)}(\phi_1(\eta),\eta) \;, \nonumber
\end{eqnarray}
we may factor $\mu$ out of the constant term in the system, i.e.
\begin{displaymath}
f^{(L)}(0;\mu,\eta) = f^{(R)}(0;\mu,\eta) = \mu b(\mu,\eta) + o(k) \;,
\end{displaymath}
where $b$ is $C^{k-1}$.
Notice the transformation does not alter the switching manifold. 
The system is now
\begin{equation}
\dot{x} = \left\{ \begin{array}{lc} 
f^{(L)}(x;\mu,\eta), & x_1 \le 0 \\
f^{(R)}(x;\mu,\eta), & x_1 \ge 0
\end{array} \right. \;,
\label{eq:theFlow}
\end{equation}
with
\begin{equation}
f^{(i)}(x;\mu,\eta) = \mu b(\mu,\eta) + A_i(\mu,\eta) x + O(|x|^2) + o(k) \;,
\label{eq:fiForm}
\end{equation}
where $A_L$ and $A_R$ are $N \times N$ matrices that are $C^{k-1}$
functions of $\mu$ and $\eta$.

Since (\ref{eq:theFlow}) is continuous,
the matrices $A_L$ and $A_R$ have matching elements
in all but possibly their first columns.
It directly follows that the {\em adjugate} matrices
(if $A$ is non-singular, then ${\rm adj}(A) \equiv \det(A) A^{-1}$)
of $A_L$ and $A_R$ share the same first row
\begin{equation}
\xi^{\sf T} = e_1^{\sf T} {\rm adj}(A_L)
= e_1^{\sf T}{\rm adj}(A_R) \;.
\label{eq:xiDef}
\end{equation}
To understand the role of the vector, $\xi$,
consider equilibria
of (\ref{eq:theFlow}).
When $\mu = \eta = 0$, the origin is an equilibrium.
For small non-zero $\mu$
each half-system has an equilibrium, $x^{*(i)}$, with first component
\begin{displaymath}
x_1^{*(i)}(\mu,\eta) = -\frac{\xi^{\sf T} b}
{\det(A_i)} \Bigg|_{\mu = \eta = 0} \mu + O(2) \;,
\end{displaymath}
provided that $\det(A_i(0,0)) \ne 0$.
Notice that our non-degeneracy assumption (\ref{eq:transvCond}),
is satisfied if $\xi^{\sf T}(0,0) b(0,0) \ne 0$.
If $x_1^{*(R)} \ge 0$, then $x^{*(R)}$ is an equilibrium of
the piecewise-smooth system (\ref{eq:theFlow}) and is said
to be {\em admissible}, otherwise it is {\em virtual}.
Similarly $x^{*(L)}$ is admissible if and only if $x_1^{*(L)} \le 0$.
Finally, notice if $\det(A_L(0,0))$ and $\det(A_R(0,0))$ are of
the same sign, then $x^{*(L)}$ and $x^{*(R)}$ are admissible for
different signs of $\mu$, whereas if $\det(A_L(0,0))$ and $\det(A_R(0,0))$
have opposite sign, $x^{*(L)}$ and $x^{*(R)}$ are admissible
for the same sign of $\mu$.
The former case is known as {\em persistence};
the latter is known as a {\em non-smooth fold} \cite{DiBu08}.

The following theorem describes dynamical phenomena
near $\mu = \eta = 0$ when $\det(A_L(0,0)) = 0$.

\addtocounter{lemma}{1}
\begin{theorem}~\\
Consider the system (\ref{eq:theFlow})
with (\ref{eq:fiForm}) and assume that $N \ge 1$ and $k \ge 4$.
Suppose that near $(\mu,\eta) = (0,0)$,
$A_L(\mu,\eta)$ has an eigenvalue $\lambda(\mu,\eta) \in \mathbb{R}$
with the associated eigenvector, $v(\mu,\eta)$. In addition, suppose
\begin{enumerate}
\renewcommand{\labelenumi}{\roman{enumi})}
\item $\lambda(0,0) = 0$ is of algebraic multiplicity 1
and is the only eigenvalue of $A_L(0,0)$ with zero real part.
\end{enumerate}
Then $v(0,0)$ has a non-zero first component
and the magnitude of $v(0,0)$ may be scaled such that,
$\xi(0,0)^{\sf T} v(0,0) = 1$. Finally suppose that
\begin{enumerate}
\renewcommand{\labelenumi}{\roman{enumi})}
\addtocounter{enumi}{1}
\item $\frac{\partial \lambda}{\partial \eta}(0,0) \ne 0$,
\item $a_0 = \xi^{\sf T}((D_x^2 f^{(L)})(v,v)) \big|_{(0,0,0)} \ne 0$,
\item $\xi(0,0)^{\sf T} b(0,0) \ne 0$.
\end{enumerate}
Then there exists a unique $C^{k-2}$ function $h : \mathbb{R} \to \mathbb{R}$
with $h(0) = h'(0) = 0$ and
\begin{equation}
h''(0) = \frac{(\frac{\partial \lambda}{\partial \eta})^2}
{2 a_0 \xi^{\sf T} b} \bigg|_{(0,0)} \;,
\end{equation}
such that in a neighborhood of $(\mu,\eta) = (0,0)$,
the curve $\mu = h(\eta)$ corresponds to a locus of saddle-node bifurcations
of equilibria of (\ref{eq:theFlow}) that are
admissible when
\begin{equation}
{\rm sgn}(\eta) = {\rm sgn}(a_0 v_1
{\textstyle \frac{\partial \lambda}{\partial \eta}}) \big|_{(0,0)} \;.
\end{equation}
\label{th:c2sn}
\end{theorem}
A proof of Theorem \ref{th:c2sn} is given in Appendix \ref{sec:PROOFS}.
The theorem implies a bifurcation diagram
like that depicted in Fig.~\ref{fig:bsSch}A.
In particular the curve of saddle-node bifurcations is tangent
to the $\eta$-axis as shown.

The second theorem, Theorem \ref{th:c2hb},
describes dynamical phenomena
near $\mu = \eta = 0$ when $A_L(0,0)$ has a
purely imaginary complex eigenvalue pair.
The method of proof is essentially a standard dimension
reduction by restriction to
the center manifold of the left-half-system
The dynamics of the resulting planar system are determined by
the following two lemmas.
Lemma \ref{le:c2hb2dgen} provides a transformation 
of the planar system to observer canonical form \cite{DiBu08}.
Lemma \ref{le:c2hb2d} describes local dynamics
of the planar system in this canonical form.

\addtocounter{theorem}{1}
\begin{lemma}~\\
Consider the two-dimensional $C^k$ ($k \ge 5$) system
\begin{equation}
\left[ \begin{array}{c} \dot{x} \\ \dot{y} \end{array} \right] =
\left[ \begin{array}{c} f(x,y;\mu,\eta) \\ g(x,y;\mu,\eta) \end{array} \right] =
\left[ \begin{array}{c} 0 \\ -\mu \end{array} \right] +
\left[ \begin{array}{cc} \eta & 1 \\ -\delta(\mu,\eta) & 0 \end{array} \right]
\left[ \begin{array}{c} x \\ y \end{array} \right] + O(|x,y|^2) + o(k) \;.
\label{eq:2dflowHB}
\end{equation}
Suppose,
\begin{enumerate}
\renewcommand{\labelenumi}{\roman{enumi})}
\item $\delta(0,0) = \omega^2$ for $\omega > 0$,
\item $a_0 \ne 0$, where
\end{enumerate}
\vspace{-5mm}
\begin{eqnarray}
a_0 & = & \frac{1}{16}(f_{xxx} + g_{xxy} + \omega^2 f_{xyy} + \omega^2 g_{yyy}) -
\frac{1}{16} f_{xy}(f_{xx} + \omega^2 f_{yy}) \nonumber \\
& & +~\frac{1}{16} g_{xy}(\frac{1}{\omega^2} g_{xx} + g_{yy}) +
\frac{1}{16}(\frac{1}{\omega^2}f_{xx}g_{xx} - \omega^2 f_{yy}g_{yy})
\label{eq:a0HB}
\end{eqnarray}
evaluated at $(x,y;\mu,\eta) = (0,0;0,0)$.
Then, near $(x,y;\mu,\eta) = (0,0;0,0)$,
there is a unique equilibrium given by $C^k$ functions
\begin{eqnarray}
x^*(\mu,\eta) & = & -\frac{1}{\omega^2} \mu + O(2) \;, \nonumber \\
y^*(\mu,\eta) & = & O(2) \;, \nonumber
\end{eqnarray}
and there exist $C^{k-1}$, $C^{k-2}$ functions $h_1,h_2 : \mathbb{R} \to \mathbb{R}$
respectively, with
\begin{eqnarray}
h_1(\mu) & = & \frac{1}{\omega^2}(f_{xx} + g_{xy}) \bigg|_{(0,0;0,0)} \mu + O(\mu^2) \;, \\
h_2(\mu) & = & h_1(\mu) - \frac{2 a_0}{\omega^4} \mu^2 + O(\mu^3) \;,
\end{eqnarray}
such that for small $\mu$,
the curve $\eta = h_1(\mu)$ corresponds to Andronov-Hopf bifurcations of $(x^*,y^*)^{\sf T}$
and the curve $\eta = h_2(\mu)$ corresponds to associated Hopf cycles intersecting the
$y$-axis tangentially at one point.
The Hopf bifurcations are supercritical if $a_0 < 0$ and subcritical if $a_0 > 0$.
The Hopf cycle lies entirely in the left-half-plane if and only if $\mu > 0$ and
$\eta$ lies between $h_1(\mu)$ and $h_2(\mu)$.
\label{le:c2hb2d}
\end{lemma}

\addtocounter{theorem}{1}
\begin{lemma}~\\
Consider the system (\ref{eq:theFlow})
with (\ref{eq:fiForm}) and assume that $N = 2$ and $k \ge 5$.
Suppose $A_L(\mu,\eta)$ has complex eigenvalues
$\lambda_\pm = \nu \pm {\rm i} \omega$ with
\begin{enumerate}
\renewcommand{\labelenumi}{\roman{enumi})}
\item $\nu(0,0) = 0$ and $\omega(0,0) > 0$,
\item $\frac{\partial \nu}{\partial \eta}(0,0) \ne 0$,
\item $\xi^{\sf T}(0,0) b(0,0) \ne 0$.
\end{enumerate}
Then there is a nonlinear transformation not altering the switching manifold
such that the left-half-flow of (\ref{eq:theFlow}) is given by (\ref{eq:2dflowHB}).
All conditions in Lemma \ref{le:c2hb2d} will be satisfied except possibly
the non-degeneracy condition, $a_0 \ne 0$.
\label{le:c2hb2dgen}
\end{lemma}

\begin{theorem}~\\
Consider the system (\ref{eq:theFlow})
with (\ref{eq:fiForm}) and assume that $N > 2$ and $k \ge 6$.
Suppose near $(\mu,\eta) = (0,0)$,
$A_L(\mu,\eta)$ has complex eigenvalues
$\lambda_\pm = \nu \pm {\rm i} \omega$
with associated eigenvectors,
$z_\pm = u^{(1)} \pm {\rm i} u^{(2)}$.
Suppose
\begin{enumerate}
\renewcommand{\labelenumi}{\roman{enumi})}
\item $\nu(0,0) = 0$, $\omega(0,0) > 0$, and $A_L(0,0)$ has
no other eigenvalues on the imaginary axis,
\item $\frac{\partial \nu}{\partial \eta}(0,0) \ne 0$,
\item $\xi^{\sf T}(0,0) b(0,0) \ne 0$,
\item either $u_1^{(1)}$ or $u_1^{(2)}$ is non-zero.
\end{enumerate}
Then, in the extended coordinate system $(x,\mu,\eta)$, there exists a
$C^{k-1}$ four-dimensional center manifold, $W^c$,
for the left-half-system that passes through the origin and is not tangent
to the switching manifold at this point.
Furthermore, $\exists i \ne 1$,
such that in the coordinate system
\begin{displaymath}
\left[ \begin{array}{c} \hat{x}_1 \\ \hat{x}_2 \end{array} \right] =
\left[ \begin{array}{c} x_1 \\ x_i \end{array} \right] \;,
\end{displaymath}
the left-half-flow of (\ref{eq:theFlow}) restricted to
$W^c$ is given by (\ref{eq:2dflowHB}) (with ``hatted'' variables)
and all conditions in Lemma \ref{le:c2hb2dgen} will be satisfied.
\label{th:c2hb}
\end{theorem}
See Appendix \ref{sec:PROOFS} for a proof.
Theorem \ref{th:c2hb} implies that a $n$-dimensional system near a codimension-two
point with a simultaneous Hopf and discontinuous bifurcation
will have a bifurcation diagram like that shown in Fig.~\ref{fig:bsSch}B.
In particular,
the curves of grazing and Hopf bifurcations are tangent to
one another at the origin.
For this scenario in two-dimensions it is known that a curve of saddle-node
bifurcations of the Hopf cycle may exist very close
to the grazing curve \cite{SiMe08}.
We have not been able to extend this result to higher dimensions.


\section{Conclusions}
\label{sec:CONCL}
\setcounter{equation}{0}

In this paper we investigated the onset of stable oscillations and  
more complex behavior in a model of {\em \sacc}~growth taken from  
\cite{JoKo99}. The model assumes an instantaneous switching between  
competing metabolic pathways resulting in a piecewise-smooth, 
continuous system of ODEs. In this paper we identified a variety of  
discontinuity induced bifurcations.

The model exhibits stable oscillations that arise from Andronov-Hopf  
bifurcations for intermediate values of the dilution rate, $D$; these  
have also been observed experimentally. As $D$ grows, the oscillation  
amplitude suddenly jumps to a much larger value just slightly beyond  
the Hopf bifurcation. We do not have a detailed explanation for this  
sudden amplitude change. As $D$ is increased further, the resulting  
stable orbits undergo a complex sequence of bifurcations causing their  
periods and amplitudes to decrease, until a second Hopf bifurcation  
occurs resulting again in a stable equilibrium.

For the model (\ref{eq:ALLdot}),
a change in the preferred metabolic pathway at an equilibrium results  
in the loss of differentiability for orbits in its neighborhood. The  
result is often a discontinuity induced bifurcation, and we have  
identified discontinuous saddle-node and  Hopf bifurcations. The  
system also exhibits codimension-two bifurcations that correspond to  
simultaneous discontinuous saddle-node and Hopf bifurcations. We have  
provided a rigorous unfolding of these scenarios from a general  
viewpoint.

While the behavior that we studied are specific to piecewise smooth,  
continuous models, a model in which this simplification is relaxed  
should still have much the same behavior. For example if the relative  
strength of two competing pathways reverses, then exponential growth  
will lead to the dominance of one over the other over a small  
parameter range and a short timescale.
Though the bifurcations generic to smooth systems are  
restricted relative to those of discontinuous systems, a rapid  
sequence of these bifurcations over a small range of parameters may  
lead to the same behavior on a rougher scale as the discontinuous  
one. This can be seen, for example, in the simplest models such as a  
smoothed one-dimensional tent map.




\appendix
\section{Proofs}
\label{sec:PROOFS}
\setcounter{equation}{0}

\noindent
{\bf Proof of Theorem \ref{th:c2sn}}\\
Recall, for any $N \times N$ matrix, $A$,
\begin{displaymath}
{\rm adj}(A) A = \det(A) I \;.
\end{displaymath}
By putting $A = A_L(0,0)$, multiplying on the left by
$e_1^{\sf T}$ and using (\ref{eq:xiDef}) we obtain
\begin{displaymath}
\xi^{\sf T}(0,0) A_L(0,0) = 0 \;.
\end{displaymath}
Thus $\xi^{\sf T}(0,0)$ is the left eigenvector of $A_L(0,0)$
for $\lambda(0,0) = 0$.
Consequently we may indeed choose the length of
the right eigenvector
$v(0,0)$ such that
$\xi^{\sf T}(0,0) v(0,0) = 1$.

Now we show $v(0,0)$ has a non-zero first element.
Suppose for a contradiction,
$v_1(0,0) = 0$.
Let $B_{ij}$ denote the $(N-1) \times (N-1)$ matrix formed by removing
the $i^{\rm th}$ row and $j^{\rm th}$ column from $A_L(0,0)$.
Let $\tilde{v} = ( v_2(0,0)~\ldots~v_N(0,0) )^{\sf T} \in \mathbb{R}^{N-1}$.
Then $\tilde{v} \ne 0$ and $B_{i1} \tilde{v} = 0$ for each $i$.
Therefore each $\det(B_{i1}) = 0$,
i.e., each element in the first column of the
cofactor matrix of $A_L(0,0)$ is zero.
Thus $\xi^{\sf T}(0,0)
= 0$
which is a contradiction.
Therefore
\begin{displaymath}
v_1(0,0) \ne 0 \;.
\end{displaymath}
Let $v^{(1)}, \ldots, v^{(N)}$ be $N$ generalized eigenvectors of $A_L(0,0)$
that form a basis of $\mathbb{R}^N$ with $v^{(1)} = v(0,0)$.
Let $V = \left[ v^{(1)}~\cdots~v^{(N)} \right]$.
We introduce the linear change of coordinates
\begin{equation}
\hat{x} = V^{-1} x \;.
\end{equation}
Let
\begin{equation}
F = \left[ \begin{array}{c}
\dot{\hat{x}} \\ \dot{\mu} \\ \dot{\eta}
\end{array} \right] = 
\left[ \begin{array}{c}
V^{-1} f^L(V \hat{x};\mu,\eta) \\ 0 \\ 0
\end{array} \right] \;,
\label{eq:Fproof}
\end{equation}
denote the $(N+2)$-dimensional $C^k$ extended left-half-flow
in the basis of generalized eigenvectors.
The Jacobian
\begin{equation}
D F(0;0,0) = \left[ \begin{array}{cccccc}
& V^{-1} A_L(0,0) V & \Bigg| & V^{-1} b(0,0) & \Bigg| & 0 \\
\hline
& 0 & \Bigg| & 0 & \Bigg| & 0
\end{array} \right] \;,
\label{eq:DFproof}
\end{equation}
has a three-dimensional nullspace, $E^c$.
The $(N+2)$-dimensional vectors
\begin{displaymath}
\{ n_1, n_2, n_3 \} =
\{ e_1,
\left[ \begin{array}{c}
\varphi \\ 1 \\ 0
\end{array} \right], e_{N+2} \} \;,
\end{displaymath}
where $n_2$ is a generalized eigenvector
and $\varphi \in \mathbb{R}^N$,
span $E^c$.
The local center manifold, $W^c$, is tangent to $E^c$, thus on $W^c$,
\begin{displaymath}
\hat{x} = H(\hat{x}_1;\mu,\eta) =
\hat{x}_1 e_1 - \mu \zeta + O(2) \;,
\end{displaymath}
where $\zeta \in \mathbb{R}^{N}$
is equal to $\varphi$ except that its first element is zero.
Notice $\xi^{\sf T}(0,0) V = e_1^{\sf T}$ thus
\begin{displaymath}
\hat{x}_1 = e_1^{\sf T} \hat{x} =
\xi^{\sf T}(0,0) V \hat{x} = \xi^{\sf T}(0,0) x \;.
\end{displaymath}
Restricted to $W^c$ the dynamics (\ref{eq:Fproof}) become the
$C^{k-1}$ system
\begin{eqnarray}
\dot{\hat{x}}_1 & = & \mu \xi^{\sf T}(0,0) b(\mu,\eta) +
\xi^{\sf T}(0,0) A_L(\mu,\eta) V H(\hat{x}_1;\mu,\eta) \nonumber \\
& & +~\xi^{\sf T}(0,0) g^L(V H(\hat{x}_1;\mu,\eta);\mu,\eta) \;,
\label{eq:x1hatdot}
\end{eqnarray}
where $g^L(x;\mu,\eta)$ denotes all
terms of $f^L$ that are nonlinear in $x$.
By expanding each term in (\ref{eq:x1hatdot})
to second order we obtain
\begin{equation}
\dot{\hat{x}}_1(\hat{x}_1;\mu,\eta) = c_1 \mu + c_2 \hat{x}_1^2 +
c_3 \hat{x}_1 \eta + c_4 \hat{x}_1 \mu + c_5 \mu^2 + c_6 \mu \eta + O(3) \;,
\label{eq:x1hatdot2}
\end{equation}
where, in particular
\begin{eqnarray}
c_1 & = & \xi(0,0)^{\sf T} b(0,0) \;, \nonumber \\
c_2 & = & \frac{a_0}{2} \;, \nonumber \\
c_3 & = & \frac{\partial \lambda}{\partial \eta}(0,0) \;. \nonumber
\end{eqnarray}
Let $\hat{x}^*_1$ be an equilibrium of (\ref{eq:x1hatdot2}).
Since $c_1 \ne 0$ by hypothesis, by the implicit function theorem
there exists a unique $C^{k-1}$ function,
$\mu_{\rm eq}(\hat{x}^*_1,\eta)$ such that
$\dot{\hat{x}}_1(\hat{x}^*_1;
\linebreak																			
\mu_{\rm eq}(\hat{x}^*_1,\eta),\eta) = 0$ and
\begin{displaymath}
\mu_{\rm eq}(\hat{x}^*_1,\eta) = -\frac{c_2}{c_1} \hat{x}^{*^2}_1 -
\frac{c_3}{c_1} \hat{x}^*_1 \eta + O(3) \;.
\end{displaymath}
Near $(\mu,\eta) = (0,0)$,
the linearization about the equilibrium $\hat{x}^*_1$, has an
associated eigenvalue of $0$ exactly when
\begin{equation}
0 = \frac{\partial \dot{\hat{x}}_1} 
{\partial \hat{x}_1} (\hat{x}^*_1;\mu_{\rm eq}(\hat{x}^*_1,\eta),\eta)
= 2 c_2 \hat{x}^*_1 + c_3 \eta + O(2) \;.
\label{eq:derivXhatdot}
\end{equation}
Since $a_0 \ne 0$ by hypothesis,
the implicit function theorem again implies
there exists a unique $C^{k-2}$ function,
$\hat{h} : \mathbb{R} \to \mathbb{R}$ such that,
(\ref{eq:derivXhatdot}) is satisfied when $\hat{x}^*_1 = \hat{h}(\eta)$ and
\begin{equation}
\hat{h}(\eta) = -\frac{c_3}{2 c_2} \eta + O(\eta^2) \;.
\label{eq:hhat}
\end{equation}
The function
\begin{equation}
h(\eta) = \mu_{\rm eq}(\hat{h}(\eta),\eta) =
\frac{c_3^2}{4 c_1 c_2} \eta^2 + O(\eta^3) \;, 
\end{equation}
is $C^{k-2}$.
We now show saddle-node bifurcations occur for the left-half-flow
on the curve $\mu = h(\eta)$ when $\eta$ is small by verifying the three
conditions of the saddle-node bifurcation theorem,
see for instance \cite{GuHo86}:
\begin{enumerate}
\renewcommand{\labelenumi}{\roman{enumi})}
\item by construction,
$D_x f^L(x^*;h(\eta),\eta)$ has a zero eigenvalue of algebraic multiplicity $1$,
and there are no other eigenvalues with zero real part
when $\eta$ is sufficiently small,
\item $w(\mu,\eta) \frac{\partial f^L}{\partial \mu}(x^*;h(\eta),\eta) =
\xi^{\sf T}(0,0) b(0,0) + O(\eta) \ne 0$ where $w(\mu,\eta)$
is the left eigenvector of $A_L(\mu,\eta)$ for $\lambda(\mu,\eta)$,
\item $w(\mu,\eta) (D_x^2 f^L(x^*;h(\eta),\eta)(v(\mu,\eta),v(\mu,\eta))) =
a_0 + O(\eta) \ne 0$.
\end{enumerate}
Finally, notice that on $W^c$
\begin{displaymath}
x_1 = e_1^{\sf T} V H(\hat{x}_1;\mu,\eta) =
v_1(0,0) \hat{x}_1 + O(2) \ne 0 \;,
\end{displaymath}
and by (\ref{eq:hhat}), when $\mu = h(\eta)$
\begin{displaymath}
x^*_1 = -\frac{c_3 v_1(0,0)}{2 c_2} \eta + O(\eta^2) \;.
\end{displaymath}
Thus the equilibrium at the saddle-node bifurcation
is admissible when
${\rm sgn}(\eta) = {\rm sgn}(c_2 c_3 v_1(0,0))
= {\rm sgn}(a_0 v_1 \frac{\partial \lambda}{\partial \eta}) \bigg|_{(0,0)}$.\\
$\Box$

\noindent
{\bf Proof of Theorem \ref{th:c2hb}}\\
Since the real and imaginary parts of the eigenvectors $z_\pm$,
$u^{(1)}$ and $u^{(2)}$, are linearly independent,
there exists an $i \ne 1$ such that
\begin{displaymath}
U = \left[ \begin{array}{cc}
u^{(1)}_1 & u^{(2)}_1 \\
u^{(1)}_i & u^{(2)}_i
\end{array} \right] \;,
\end{displaymath}
is non-singular.
For the remainder of this proof we will set $i = 2$, w.l.o.g.
Define two new vectors by
\begin{equation}
\left[ v^{(1)}~v^{(2)} \right] =
\left[ u^{(1)}~u^{(2)} \right] U^{-1} \;,
\label{eq:v1v2Def}
\end{equation}
let
\begin{displaymath}
V = \left[ v^{(1)}~v^{(2)}~e_3~\cdots~e_N \right] \;,
\end{displaymath}
and introduce the new coordinate system
\begin{equation}
\hat{x} = V^{-1} x \;.
\end{equation}
Note the inclusion of the matrix, $U^{-1}$, in (\ref{eq:v1v2Def})
allows for simplification below, in particular,
$e_i^{\sf T} V = e_i^{\sf T} V^{-1} = e_i^{\sf T}$
for $i = 1,2$.
The $(N+2)$-dimensional $C^k$ extended left-half-flow
in the new coordinates is given by (\ref{eq:Fproof}) as before.
The Jacobian, $DF(0;0,0)$, (\ref{eq:DFproof}),
has a four-dimensional linear center manifold, $E^c$,
spanned by
\begin{displaymath}
\{ n_1, n_2, n_3, n_4 \} =
\{ e_1, e_2,
\left[ \begin{array}{c}
-(A_L(0,0) V)^{-1} b \\ 1 \\ 0
\end{array} \right], e_{N+2} \} \;.
\end{displaymath}
Notice $W^c$ is not tangent to the switching manifold by condition (iv) of the theorem.
On the local center manifold
\begin{displaymath}
\hat{x} = H(\hat{x}_1,\hat{x}_2;\mu,\eta) =
\hat{x}_1 e_1 + \hat{x}_2 e_2 - \zeta \mu + O(2) \;,
\end{displaymath}
where $\zeta \in \mathbb{R}^{N}$
is equal to $(A_L(0,0) V)^{-1} b$
except that its first two elements are zero.
The dynamics on $W^c$ are described by
\begin{eqnarray}
\left[ \begin{array}{c} \dot{\hat{x}}_1 \\ \dot{\hat{x}}_2 \end{array} \right] & = &
\left[ \begin{array}{c} e_1^{\sf T} \\
e_2^{\sf T} \end{array} \right] \bigg( V^{-1} \mu b 
+ V^{-1} A_L V H(\hat{x}_1,\hat{x}_2;\mu,\eta) \nonumber \\
& & +~V^{-1} g^{(L)}(V H(\hat{x}_1,\hat{x}_2;\mu,\eta);\mu,\eta) \bigg) \;, \nonumber
\end{eqnarray}
where $g^{(L)}$ represents all terms of $f^{(L)}$ that are nonlinear in $x$.
By using (\ref{eq:v1v2Def}) and
\begin{displaymath}
A_L \left[ u^{(1)}~u^{(2)} \right] = \left[ u^{(1)}~u^{(2)} \right] D \;,
\end{displaymath}
where
\begin{displaymath}
D = \left[ \begin{array}{cc}
\nu & \omega \\
-\omega & \nu
\end{array} \right] \;,
\end{displaymath}
we obtain
\begin{equation}
\left[ \begin{array}{c} \dot{\hat{x}}_1 \\ \dot{\hat{x}}_2 \end{array} \right] =
\mu \hat{b}(\mu,\eta) + \hat{A}_L(\mu,\eta)
\left[ \begin{array}{c} \hat{x}_1 \\ \hat{x}_2 \end{array} \right] + O(2) \;, 
\end{equation}
where
\begin{displaymath}
\hat{A}_L =
\left[ \begin{array}{c} e_1^{\sf T} \\ e_2^{\sf T} \end{array} \right]
V^{-1} A_L V
\left[ e_1~e_2 \right] =
U D U^{-1} \;,
\end{displaymath}
and
\begin{eqnarray*}
\hat{b} & = & 
\left[ \begin{array}{c} e_1^{\sf T} \\ e_2^{\sf T} \end{array} \right] V^{-1} b -
\left[ \begin{array}{c} e_1^{\sf T} \\ e_2^{\sf T} \end{array} \right] V^{-1} A_L V \zeta =
\left[ \begin{array}{c} e_1^{\sf T} \\ e_2^{\sf T} \end{array} \right] V^{-1} A_L
\left[ v^{(1)}~v^{(2)}~0 \cdots 0 \right] A_L^{-1} b \\
& = & \left[ \hat{A}_L~0 \cdots 0 \right] A_L^{-1} b \;.
\end{eqnarray*}
Finally, it is easily verified that
\begin{displaymath}
\hat{\xi}^{\sf T} \hat{b} =
\frac{\det(\hat{A}_L)}{\det(A_L)} \xi^{\sf T} b \ne 0 \;,
\end{displaymath}
where $\hat{\xi}^{\sf T} = e_1^{\sf T} {\rm adj}(\hat{A}_L)$.\\
$\Box$

\section{Functions and Parameter Values}
\label{sec:PARAM}
\setcounter{equation}{0}

The following is a complete list of functions that are present in the model
\begin{eqnarray*}
\mu_i & = & \mu_{i,{\rm max}} \frac{\mu_{i,{\rm max}}
+ \beta}{\alpha + \alpha^*} \;, ~~~~~i = 1,2,3 \;, \\
r_1 & = & \mu_1 e_1 \frac{G}{K_1 + G} \;, \\
r_2 & = & \mu_2 e_2 \frac{E}{K_2 + E} \frac{O}{K_{{\rm O}_2}+O} \;, \\
r_3 & = & \mu_3 e_3 \frac{G}{K_3 + G} \frac{O}{K_{{\rm O}_3}+O} \;, \\
u_i & = & \frac{r_i}{\sum_j r_j},~~~~~i = 1,2,3 \;, \\
v_i & = & \frac{r_i}{\max_j r_j},~~~~~i = 1,2,3 \;.
\end{eqnarray*}

\begin{table}[h]
\begin{center}
\begin{tabular}{l@{\hspace{10mm}}l@{\hspace{5mm}}|@{\hspace{5mm}}l@{\hspace{10mm}}l}
$G_0$ & 10 gL$^{-1}$ & $K_1$ & 0.05 gL$^{-1}$ \\
$Y_1$ & 0.16 gg$^{-1}$ & $K_2$ & 0.01 gL$^{-1}$ \\
$Y_2$ & 0.75 gg$^{-1}$ & $K_3$ & 0.001 gL$^{-1}$ \\
$Y_3$ & 0.60 gg$^{-1}$ & $K_{{\rm O}_2}$ & 0.01 mgL$^{-1}$ \\
$\phi_1$ & 0.403 & $K_{{\rm O}_3}$ & 2.2 mgL$^{-1}$ \\
$\phi_2$ & 2000 mgg$^{-1}$ & $\gamma_1$ & 10 \\
$\phi_3$ & 1000 mgg$^{-1}$ & $\gamma_2$ & 10 \\
$\phi_4$ & 0.95 & $\gamma_3$ & 0.8 gg$^{-1}$ \\
$O^*$ & 7.5 mgL$^{-1}$ & $\mu_{1,{\rm max}}$ & 0.44 h$^{-1}$ \\
$\alpha$ & 0.3 gg$^{-1}$h$^{-1}$ & $\mu_{2,{\rm max}}$ & 0.19 h$^{-1}$ \\
$\alpha^*$ & 0.1 gg$^{-1}$h$^{-1}$ & $\mu_{3,{\rm max}}$ & 0.36 h$^{-1}$ \\
$\beta$ & 0.7 h$^{-1}$ & & \\
\end{tabular}
\end{center}
\caption{Parameter values used for all numerical investigations
in this paper.}
\end{table}


\end{document}